\documentclass[fleqn, usenatbib]{mnras}

\usepackage{newtxtext}
\usepackage{mathptmx}

\usepackage{ae,aecompl}

\usepackage{graphicx}    
\usepackage{amsmath}    
\usepackage{amssymb}    
\usepackage{fourier}
\usepackage{romannum}
\usepackage{enumerate}
\usepackage{enumitem}
\usepackage[normalem]{ulem}
\usepackage{ragged2e}
\usepackage{xcolor}
\usepackage{color}
\usepackage{natbib}

\title[DM's influence on MBHB in Leo I]{Dark Matter's influence on Evolution of MBHB in Dwarf Galaxies: A Case Study of Leo I dSph}

\author[Junaid et. al]{Muhammad Junaid$^{1,2}$\thanks{E-mail: mjunaidtariq94@gmail.com}\\
    $^{1}$Department of Space Science, Institute of Space Technology, Islamabad 44000, Pakistan\\
    $^{2}$Space and Astrophysics Research Lab (SARL), National Centre of GIS and Space Applications (NCGSA), Islamabad 44000, Pakistan\\
}
\date{Accepted XXX. Received YYY; in original form ZZZ}

\pubyear{2026}

\begin{document}
    \label{firstpage}
    \pagerange{\pageref{firstpage}--\pageref{lastpage}}
    \maketitle
    \begin{abstract}

        In this study, we investigate the dynamical evolution of a massive binary black hole (MBHB) in the Leo I dwarf spheroidal galaxy model and examine how dark matter along with stellar matter's gravitational interactions influence its long-term behavior.
        Using high-resolution direct N-body simulations, we follow the orbital evolution of the binary within a realistic model of the Leo I stellar and dark matter distribution.
        We found that the binary separation decreases from an initial 300-parsec orbit to roughly 1 parsec over a period of about 2 Gyr, primarily driven by dynamical friction and stellar hardening.
        The orbital evolution then stalls at this scale, illustrating the well-known ``final parsec problem''.
        During this phase, the binary also develops increasing orbital eccentricity and produces a modest redistribution of the inner mass profiles in some cases.
        We then further estimate the final stage of the system’s evolution using gravitational-wave emission models and find that the binary is unlikely to merge within a Hubble time.
        The prolonged dynamical friction phase appears to be related to the low stellar and dark matter densities in Leo I. These results suggest that massive binary black holes in dwarf spheroidal galaxies such as Leo I will not contribute to the gravitational-waves detectable from LISA even if dark matter is considered.

    \end{abstract}




    \section{Introduction}
    \label{sec-intro}

    \subsection{Leo~I}

    Leo~I is a dwarf spheroidal (dSph) galaxy and a well-studied member
    of the Local Group, gravitationally associated with the Milky Way
    subsystem.
    Early investigations of its stellar kinematics and
    metallicities established it as a chemically and dynamically distinct
    system~\cite{Koch2007}, while studies of its chemical abundances
    provided further constraints on its stellar populations and
    evolutionary history~\cite{Bosler2007}.
    Structural analyses have
    revealed complex, multi-age stellar populations~\cite{Okamoto2011},
    and Leo~I has been classified as an unquenched, very low-mass galaxy,
    suggesting an evolutionary path distinct from other dwarf galaxies~\cite{McQuinn2015}.
    Its dark matter halo profile is consistent with
    a central core, with a total mass comparable to other Milky Way dSph
    satellites~\cite{Koch2007, Mateo1998}.
    Broader studies have also
    placed Leo~I within the context of dSph progenitor populations and
    the evolutionary puzzle of the Local Group~\cite{Grebel2003}, while
    its kinematics have been used to probe dark matter halo structure at
    the low-mass end of the galaxy luminosity function~\cite{Salucci2012}.

    A particularly significant finding emerged from integrated-light
    measurements and previously published stellar velocity data, which
    revealed a consistent increase in velocity dispersion toward the
    center of Leo~I, with an integrated-light value of $11.76 \pm 0.66$
    km/s within $75''$~\cite{bustamante2021}.
    Axisymmetric, orbit-based
    dynamical models accounting for the stellar mass-to-light ratio, dark
    matter halo structure, and tidal effects from the Milky Way at large
    radii consistently indicate a central black hole mass of
    $M_{\bullet} \sim 3.3 \pm 2 \times 10^6~M_{\odot}$~\cite{bustamante2021}.
    The absence of a black hole is ruled out at greater than $95\%$
    confidence.
    This makes Leo~I the first dSph galaxy in which a central
    black hole has been detected using spatially resolved kinematics, with
    a mass comparable to both the total stellar mass of the system and the
    Milky Way's central black hole~\cite{bustamante2021}.

    The detected black hole mass places Leo~I well off the standard
    $M_{\bullet}$--$M_{\star}$ scaling relation by more than two orders
    of magnitude~\cite{pacucci2023extreme}.
    One proposed explanation is
    that Leo~I's stellar system was originally more massive and consistent
    with the scaling relation, but lost a substantial fraction of its
    stellar mass through extreme tidal stripping during close passages
    within the Milky Way's virial radius~\cite{pacucci2023extreme}.
    Analytical models and N-body simulations estimate a stellar mass loss
    of $32\%$--$78\%$ depending on pericenter distance and velocity
    dispersion, with up to $90\%$ possible with additional passages~\cite{bustamante2021, pacucci2023extreme}.
    A tidal stream aligned
    along the line of sight toward Leo~I is noted as potential evidence
    of this stripping event, though its detection in current Gaia data
    remains challenging~\cite{pacucci2023extreme}.

    \subsection{Dark Matter}

    Dark matter is a non-luminous component that interacts gravitationally
    but does not emit or absorb electromagnetic radiation.
    Within the
    $\Lambda$CDM framework, it constitutes approximately $25\%$ of the
    total mass-energy content of the universe.
    Its existence was first
    postulated by Fritz Zwicky in 1933 to account for the anomalous
    orbital velocities of galaxies within the Coma cluster~\cite{Berera2011}.
    Since then, a broad range of observational evidence has supported its
    presence, including galaxy rotation curves, gravitational lensing, and
    the large-scale structure of the universe~\cite{Abramowski2015, Esmaili2012}.

    The nature of dark matter remains one of the central open questions in
    modern physics.
    Leading theoretical candidates include weakly
    interacting massive particles (WIMPs), axions, and sterile neutrinos~\cite{Abe2020, Pefianco2022}.
    On cosmological scales, dark matter
    provides the gravitational scaffolding for structure formation and the
    cosmic web~\cite{Spergel2000}, while at galactic scales it governs
    rotation curves, mass distributions, and dynamical stability~\cite{Bertone2005, Kronawitter2000, Guo2009}.
    Direct detection
    experiments, space-based observatories, and high-energy colliders
    continue to search for dark matter interactions with ordinary matter.

    \subsection{Evidence for Dark Matter}

    Three key observational pillars support the existence of dark matter:

    \begin{enumerate}

        \item \textbf{Galaxy Rotation Curves:} Observations of spiral
        galaxy rotation curves, systematically pursued from the late 1970s,
        reveal flat velocity profiles at large radii inconsistent with the
        distribution of luminous matter alone.
        This flatness implies the
        presence of a significant unseen mass component extending beyond
        the visible disk.

        \item \textbf{Gravitational Lensing:} The bending and distortion
        of light from background sources by foreground mass concentrations
        provides a direct probe of total matter content independent of
        luminosity.
        Both strong and weak lensing observations have
        confirmed substantial dark matter distributions in galaxy clusters
        and at cosmological scales~\cite{Bartelmann2001, Calore2020,
            Pires2010}.

        \item \textbf{Cosmic Microwave Background (CMB):} Fluctuations in
        the CMB and their correlation with large-scale structure are
        consistent with the gravitational influence of dark matter on
        density perturbations in the early universe~\cite{Aslanyan2016}.

    \end{enumerate}

    \subsection{The Core-Cusp Problem}

    A persistent tension exists between $\Lambda$CDM predictions and
    observations of dwarf galaxy interiors.
    N-body simulations of structure formation predict that dark matter halos should follow a Navarro-Frenk-White (NFW) density profile, characterized by a central cusp where $\rho_\textrm{DM} \propto r^{-1}$.
    In contrast, observed rotation curves of gas-rich dwarf irregular galaxies favor constant-density cores in the inner halo, inconsistent with the predicted cusp.

    Hydrodynamical simulations that incorporate baryonic processes offer a resolution through what is termed cusp-core transformation.
    As gas cools and condenses at galactic centers, stellar feedback drives repeated outflows that transfer energy and angular momentum to the dark matter, gradually flattening the cuspy NFW profile into a shallower core over multiple cycles of gas inflow and outflow~\cite{read2016dark}.
    This gravitational heating mechanism, sometimes called dark matter heating, reproduces observed core profiles in low-mass galaxies within a $\Lambda$CDM framework.

    \subsection{Supermassive Black Hole Binaries}

    The closest known supermassive black hole (SMBH) binary was identified in the radio galaxy 0402+379.
    Multi-frequency Very Long Baseline Array (VLBA) observations identified two central compact flat-spectrum components~\cite{maness2004}, subsequently confirmed as two SMBHs within a single galaxy by Rodriguez et al. \cite{rodriguez2006}.
    Their projected separation is 7~pc, and the combined system mass is estimated at $1.5 \times 10^8~M_{\odot}$~\cite{maness2004, rodriguez2006}.
    Galaxy mergers are widely regarded as the primary formation channel for SMBH binaries, making their study central to understanding galaxy evolution.

    \subsection{Evolution of SMBH Binaries}

    The dynamical evolution of SMBH pairs following a galaxy merger
    proceeds through three broad phases, as outlined by~\cite{begelman1980massive}:

    \begin{enumerate}

        \item \textbf{Dynamical Friction:} The galactic cores migrate
        toward the merger center under dynamical friction from background
        stars, gas, and dark matter.
        The cores merge through violent
        relaxation, forming a new nucleus.
        A bound binary forms once the
        SMBH separation falls below the gravitational influence radius.

        \item \textbf{Binary Hardening:} The binary continues to shrink
        through dynamical friction and three-body gravitational slingshot
        interactions with stars on intersecting orbits.
        These stars carry
        away energy and angular momentum and are ejected at velocities
        comparable to the binary orbital velocity.

        \item \textbf{Gravitational Wave Inspiral:} At sufficiently small
        separations, gravitational wave emission dominates the energy and
        angular momentum loss, driving rapid coalescence of the two black
        holes.

    \end{enumerate}

    \subsection{Stellar Dynamical Hardening}

    Following the dynamical friction phase, the SMBH binary enters the
    hardening regime.
    For a binary with component masses $M_{\bullet_1}$
    and $M_{\bullet_2}$, total mass $M_{\bullet} = M_{\bullet_1} +
    M_{\bullet_2}$, mass ratio $q = M_{\bullet_2}/M_{\bullet_1}$, and
    reduced mass $\mu = M_{\bullet_1}M_{\bullet_2}/M_{\bullet}$, the
    gravitational influence radius is defined as $r_h = GM_{\bullet}/\sigma^2$,
    where $\sigma$ is the one-dimensional stellar velocity dispersion.
    Binaries are classified as hard or soft relative to the mean stellar
    kinetic energy per unit mass, following Heggie's law: hard binaries
    statistically harden further while soft binaries soften~\cite{quinlan1996dynamical}.
    The hardening rate is driven by
    three-body encounters and is characterized by the dimensionless
    hardening parameter $H$~\cite{quinlan1996dynamical}.

    \subsection{Gravitational Wave Emission}

    As the binary separation decreases sufficiently, gravitational wave
    emission takes over as the dominant energy loss mechanism.
    The
    characteristic frequency of gravitational waves is inversely
    proportional to the black hole mass, placing SMBH mergers in the
    nanohertz to millihertz frequency band.
    This is below the sensitivity
    range of ground-based detectors and requires space-based observatories
    such as LISA or Pulsar Timing Arrays (PTAs).
    Black hole mergers in
    the mass range $10^3$--$10^7~M_{\odot}$ are expected to be detectable
    by LISA with a mass precision of approximately $5\%$ \citep{bellovary_2020}.
    The orbital evolution under gravitational wave emission follows
    Peters' (1964) formulation~\cite{peters1964gravitational}, which
    describes the decay of both semi-major axis and eccentricity as
    functions of the binary parameters, with coalescence timescale
    $T_{\rm GW}$ dependent on the initial separation, eccentricity,
    and component masses.

    The manuscript is structured as follows: Section~\ref{sec:galaxies} outlines the modelling and simulation framework and tools, Section~\ref{sec:results} summarizes the data obtained, and Section~\ref{concluions} evaluates the implications of the work.

    \section{Initial Model Setup} \label{sec:galaxies}

    \subsection{Galaxy Mass Distribution Functions}

    Galaxies exhibit a wide range of structures and dynamics.
    Understanding the distribution of mass within galaxies is crucial for unraveling their formation and evolution.
    One of the key aspects in modeling the mass distribution of galaxies is the use of spherical models.
    These models provide valuable insights into the gravitational potential, dynamics, and structural properties of galaxies.
    The study of nuclear star clusters, globular clusters, massive black holes, and dark matter halo within the context of spherical systems has significantly advanced our understanding of galaxy formation and evolution.
    Galaxy models for spherical systems are crucial for understanding the structure and dynamics of galaxies.
    Various models have been developed to describe the distribution of mass in galaxies, each with its unique characteristics and applications.

    The Dehnen model is a significant spherical galaxy model that has been widely used to represent the mass distribution in galaxies.
    This model, introduced by~\cite{1993Dehnen}, is precious for studying the dynamics of spherical stellar systems.
    The Dehnen model is characterized by its flexibility in describing the density profiles of galaxies, making it applicable to a wide range of galactic systems.
    It is also known as the $\eta-Tremaine$ model.
    The density profile by Dehnen is:

    \begin{equation}\label{eq:4.4}
        \rho_\text{dehnen}(r) = \rho_o \left(\dfrac{r}{r_s}\right)^{-\gamma} \left(1 + \dfrac{r}{r_s}\right)^{(\gamma - 4)}
    \end{equation}
    here $r_s$ is the scale radius, and $\rho_0$ is the normalization density which is ${(3-\gamma)M}/{4\pi r_s^3}$.
    The Dehnen cumulative mass profile is defined by:
    \begin{equation}\label{eq4.5}
        M(r)=M\left[\frac{r}{r+r_s}\right]^{3-\gamma}
    \end{equation}

    where $M$ is the total mass of the component to be modeled and $\gamma$ is the cusp exponent which should be $0 \leq \gamma$ \textbf{<} $ 3 $.
    The models by Jaffe and Hernquist correspond to $\gamma = 2$ and $1$, respectively \citep{1993Dehnen}.
    These models have been utilized to study the gravitational potential and dynamics of spherical stellar systems, contributing to our understanding of galaxy structure and evolution.

    J. Read presented the core NFW profile in which the inner dark matter profile that can transition from cusp to core is dependent on a function, this makes the profile closer to what we deduce from observations. \cite{read2016dark}

    \subsection{AGAMA}

    AGAMA, which stands for Action-based Galaxy Modeling Architecture, is a powerful tool used to generate stable galaxy models using Schwarzschild modeling.
    This python package is designed to construct self-consistent, multi-component galaxy models in dynamic equilibrium by using the distribution function in space and solving for the virial theorem; $2K+U=0$, where $K$ is the kinetic energy and $U$ is the gravitational potential energy of the system.
    This approach allows for the construction of detailed models that accurately represent the mass distribution, kinematics, and orbital properties of stars within galaxies.
    AGAMA is unique in its ability to handle a wide range of galaxy components, including dark matter, stellar disks, bulges, and halos, allowing for the creation of complex and realistic galaxy models.
    The code uses a flexible and efficient algorithm to generate stable models that closely match observational data, making it a valuable tool for studying the formation and evolution of galaxies.
    It takes different parameters to construct a galaxy model: mass of the component or normalized density, axis ratios (p and q), density function, scale radius, slopes ($\alpha,\beta,\gamma$) or the S{\`e}rsic index (n), and cut-off radius with cut-off strength\cite{Vasiliev_2018}.

    \subsection{Model Generation}
    To model the galaxy's overall mass distribution, we used Dehnen's distribution function and used AGAMA. Our model successfully reproduced the mass profiles of Bustamnte's data.
    We fitted both its slope parameter and the masses of the NFW model on the Dehnen model for dark matter in Leo-I. We then limit the profile to approximately 300 parsecs of the galaxy's center, consistent with the results of the previous literature.
    Beyond 300 parsecs, the cumulative mass profile flattened out while the density declined sharply as we intended.
    We constrained our model to only 300 parsecs to study the high-resolution dynamics of Leo-I with respect to binary black holes.

    \begin{table}
        \centering
        \resizebox{\columnwidth}{!}{
        \begin{tabular}{c c c c c c}
            \hline
            \multicolumn{6}{c}{Modified Leo-I model} \\
            \hline
            Component      & M ($10^6 M_{\odot}$) & M (MU) & $r_s$ ($kpc$)     & N         & $\gamma$ \\
            \hline
            Black hole     & 3.3                  & 0.22   & $1\times 10^{-5}$ & 1         & 0        \\
            Stellar matter & 4.2                  & 0.28   & 0.1               & 7,17,948  & 0.6      \\
            Dark matter    & 7.5                  & 0.5    & 0.2               & 1,282,051 & 0.6      \\
            Full model     & 15                   & 1      & -                 & 2,000,000 & -        \\
            \hline
        \end{tabular}
    }
        \caption{Parameters for Leo-I model: M is mass, $r_s$ is scale radius, N is the number of particles and $\gamma$ is the Dehnen slope.}
        \label{tab:5.1}
    \end{table}

    We chose 2 million particles and gave each stellar and dark matter particle equal mass.
    The mass of a star and dark matter particle is $3.89\times10^{-7}$ in model units and 5.835 $M_{\odot}$ in physical units.
    As the dark matter's mass is greater than our model's stellar mass, the number of dark matter particles will also be higher than the number of stars.

\begin{figure}
        \centering
        \includegraphics[width=\columnwidth] {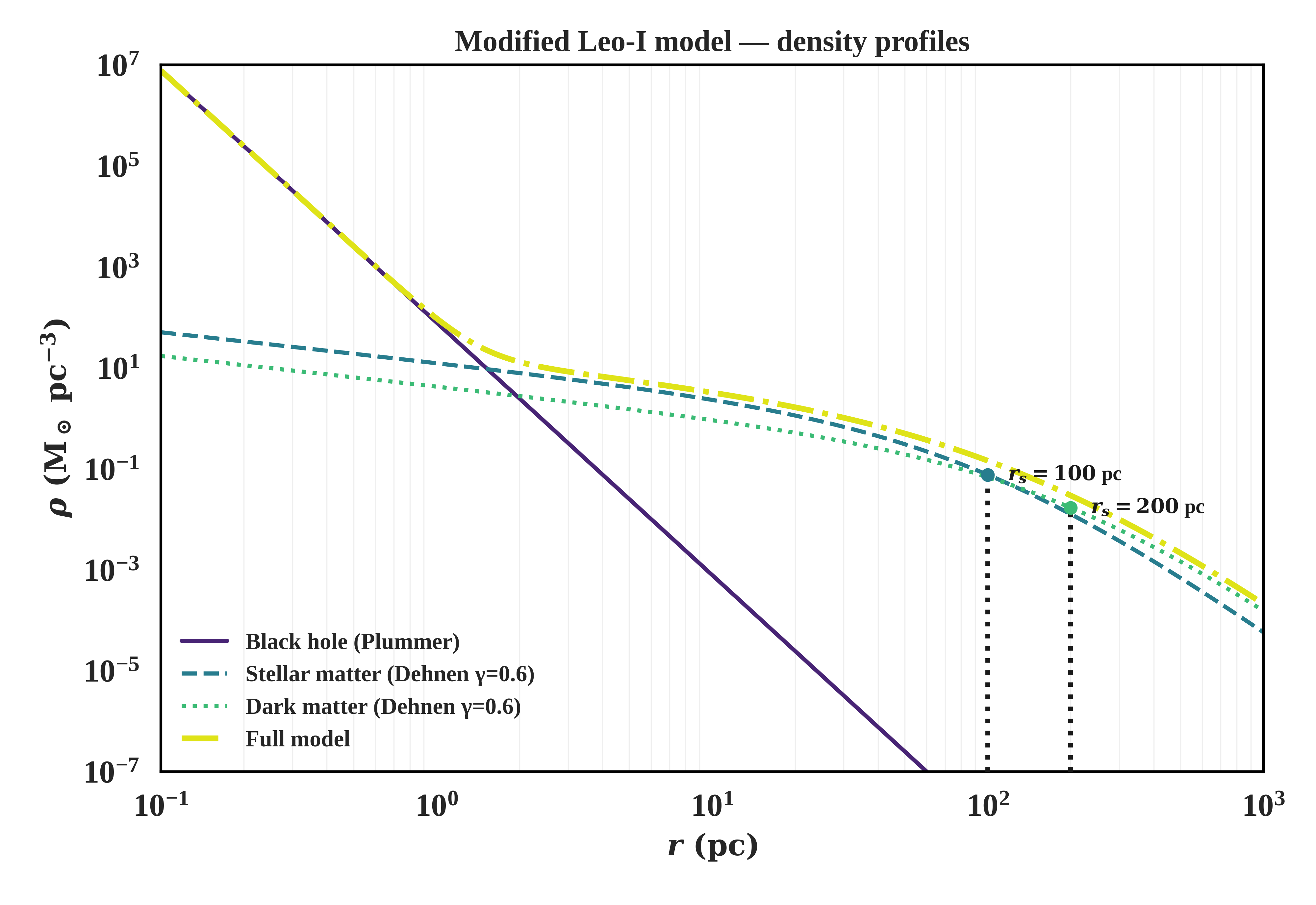}
        \caption{Density profiles of the Modified Leo I model components over the radial range (r) from 0.1 pc to 1000 pc. The black hole is modeled with a Plummer profile, while the stellar and dark matter components follow Dehnen profiles. (see Table 1 for model parameters). The full model (dash-dot line) is the sum of all three components. Dotted vertical lines indicate the scale radii of the stellar and dark matter distributions.}\label{fig:5.1}
    \end{figure}

    This approach allowed us to generate a modified model for the central region of the Leo-I dSph galaxy in equilibrium using AGAMA. The model's density and cumulative mass profile aligned with analytical analysis, testing whether its properties could be reproduced through similar assumptions about its black holes, dark matter, and stellar components as shown in \eqref{fig:5.1} for the whole model and its components.

    Since Leo-I is a dwarf spheroidal, the axis ratios of both of our components (stellar and dark matter) were chosen and shown in table~\ref{tab:5.2}.
    We also generated a visual representation of our modified Leo-I mass model using the Glnemo2 N-body visualization tool.
    Glnemo2 renders the positions of each particle.

    \begin{table}
        \centering
        \begin{tabular}{c c c}
            \hline
            Component      & $p=b/a$ & $q=c/a$ \\
            \hline
            Stellar matter & 0.9     & 0.8     \\
            Dark matter    & 0.9     & 0.8     \\
            \hline
        \end{tabular}
        \caption{Axis ratios for modified Leo-I model}
        \label{tab:5.2}
    \end{table}

    \newcommand{\mgal}{M_\text{gal}}

    \subsection{Units}
    We adopted the model units in such a way that $\mgal=G=r_o=1.$ The model units can be scaled to physical units by changing the galaxy's mass $M_\text{gal}$ and scale radius $r_o$ to physical units, when combined with the relations below we can convert time and velocity from model to physical units~\cite{gualandris2008ejection}.
    \begin{equation}
        [T]=\left(\frac{G\mgal}{{r_o}^3}\right)^{-1/2}
        = 1.5\,\text{Myr}\left(\frac{\mgal}{10^{11}M_{\odot}}\right)^{-1/2} \left(\frac{r_o}{\text{kpc}}\right)^{3/2}
    \end{equation}
    \begin{equation}
        [V]=\left(\frac{G\mgal}{r_o}\right)^{1/2}
        = 655\,\text{kms}^{-1}\left(\frac{\mgal}{10^{11}M_{\odot}}\right)^{1/2}\left(\frac{r_o}{\text{kpc}}\right)^{-1/2}
    \end{equation}

    Our model's data is in model units [MU] and their physical equivalence is given in the table below at table \eqref{tab:5.3}:

    \begin{table}
        \centering
        \begin{tabular}{c c c}
            \hline
            Quantity & Model Units & Physical Units              \\
            \hline
            Mass     & 1 [M]       & $1.5\times10^7$ $M_{\odot}$ \\
            Length   & 1 [L]       & 1 kpc                       \\
            Velocity & 1 [V]       & $7.97$ km/s               \\
            Time     & 1 [T]       & 121.7372 Myrs               \\
            \hline
        \end{tabular}
        \caption{Model units and their equivalence in physical units.}
        \label{tab:5.3}
    \end{table}

    \subsection{Secondary Black Hole}

    We inserted a secondary intermediate-mass black hole (IMBH) by modifying the second particle's mass, position, and velocity in our model.
    We chose the mass of the secondary black hole to be 100 times and 40 times less than that of our primary SMBH; hence, the mass ratios are $q_1 = 0.01$ and $q_2 = 0.025$, corresponding to secondary black hole masses of $3.3 \times 10^4 M_\odot$ and $8.25 \times 10^4 M_\odot$, respectively.
    We introduced the black hole at a distance of $r_{ini} = 300$ $pc$.
    We chose the secondary black hole's velocity as 50\% of the circular velocity at that radius for a particle this massive.
    The position was added in the x direction and velocity in the y component.
    This was deliberately done to make the initial orbit elliptical and in xy-plane.

    \begin{equation}
        v_\text{cir}=\sqrt{\frac{GM_\text{enc}(r_\text{ini})}{r_\text{ini}}}
    \end{equation}

    \subsection{N-body Code}

    N-body simulations are fundamental tools for studying the gravitational interactions and dynamics of astrophysical systems.
    These simulations involve the numerical integration of the equations of motion for a large number of particles, typically representing stars, dark matter, or other celestial objects, under the influence of gravitational forces.
    To perform these simulations, various N-body codes and integration methods have been developed, each with unique features and applications.
    In astrophysics, a system with N-particles interacting with each other gravitationally has a force $F_i$ acting on each particle with mass $m_i$:

    \begin{equation}
        \textbf{F}_i = - m_i   \sum_{j=1, j\ne i}^N \frac{m_j(\textbf{r}_i-\textbf{r}_j)}{{(|\textbf{r}_i-\textbf{r}_j|^2+\epsilon_{ij}^2)}^{3/2}}
        \label{eq:4.14}
    \end{equation}
    here,  $\epsilon$ is called the softening parameter.
    It is used to stop close body encounters and from further binaries to be formed.
    The value of softening parameter between two particles is then calculated to be:
    \begin{equation}
        \epsilon_{ij}=\sqrt{\frac{\epsilon_i^2+\epsilon_j^2}{2}}
    \end{equation}
    Gravitational forces on each particle from all the particles in the system are computed by N-body code, and for a number of time steps, the computational difficulty scales as $N^2$.

    \subsubsection{Hermite Integrator Scheme}

    The Hermite integrator is a numerical method commonly used in N-body simulations to accurately solve the equations of motion for a system of particles under the influence of gravitational forces.
    This integration method is particularly well-suited for simulating the dynamics of astrophysical systems, such as galaxies and star clusters, where the gravitational interactions between a large number of particles need to be accurately calculated over extended periods of time.
    The Hermite integrator is known for its high accuracy and efficiency in long-term gravitational interactions.
    It achieves this by using higher-order terms in the Taylor expansion of the equations of motion, allowing for more precise predictions of particle trajectories and gravitational forces.
    Additionally, the Hermite integrator can adapt its time step to account for variations in the gravitational forces, ensuring stable and accurate simulations over extended timescales.
    The Hermite fourth-order integrator is called $4^{th}-$order because the predictor term considers the third-order polynomial, and then after obtaining the acceleration, adds a $4^{th}-$order corrector term. \cite{Harfst2007}.

    \subsubsection{$\varphi-GPU$}

    The parallel processing capabilities of GPUs allow for the simultaneous execution of a large number of computational tasks, making them well-suited for accelerating N-body simulations, which involve the gravitational interactions of a vast number of particles.

    $\varphi-GPU$ is a computational framework designed for performing N-body simulations, particularly focusing on the dynamics and interactions of particles in astrophysical systems.
    It is based on $\varphi-$GRAPE6a (GRAvity PipelinE) code which can do the same function but on special boards.
    This framework leverages the computational power of Graphics Processing Units (GPUs) to speed up the simulations, enabling the efficient and high-performance modeling of complex gravitational systems.
    The $\varphi-$GPU framework harnesses the parallel processing power of GPUs to efficiently calculate the gravitational forces and trajectories of particles in N-body simulations.
    By distributing the computational workload across numerous processing cores within the GPU, $\varphi-$GPU significantly speeds up the simulation time, enabling researchers to study the dynamics and evolution of astrophysical systems with high precision and computational efficiency.
    The $\varphi-$GPU code builds upon a CPU serial N-body code, written in C++.
    It fully parallelizes the code using MPI. Additionally, it accesses GPUs directly through the NVIDIA CUDA library for native GPU support. \cite{2011Berczik}

    Before running long simulations, we performed a stability analysis of our Leo-I model using the $\varphi-GPU$ N-body solver.
    As $\varphi-GPU$ does not incorporate regularization methods for close encounters, gravitational softening was required for all particle interactions.
    A softening length of $1\times10^{-4}$ in our model units was adopted.
    This value was chosen to be sufficiently small to maximize the collisional nature of the dense stellar and dark matter components while resolving interactions down to tens of parsecs.
    The softening for the SMBH particle is equal to zero as shown in table \eqref{tab:5.4}

    \begin{table}
        \centering
        \begin{tabular}{c c}
            \hline
            \multicolumn{2}{c}{Softening length $(\epsilon)$ in kpc} \\
            \hline
            $\epsilon_\text{SMBH}$  & 0.0              \\
            $\epsilon_{\star}$ & $1\times10^{-4}$ \\
            $\epsilon_\text{DM}$    & $1\times10^{-4}$ \\
            \hline
        \end{tabular}
        \caption{Softening length for different components particles of modified Leo-I model in kpc.}
        \label{tab:5.4}
    \end{table}
    The model was evolved in the first run of the simulations for 1.5 model time units, equivalent to 180 million years.
    Density profiles and cumulative mass profiles were calculated at several timestamps throughout the simulation.
    We find that the model remains stable over this period, with no significant deviations from the initial conditions.

    \section{Results}\label{sec:results}

    The following sections present the time-resolved properties of the binary system and its impact on the host galaxy.
    We investigate the orbital evolution of a massive black hole binary (MBHB) in the central 300 pc of a Leo I for mass ratio $q = 1:100$ as well as $q = 1:40$.

    \begin{figure*}
        \centering
        \includegraphics[width=\textwidth]{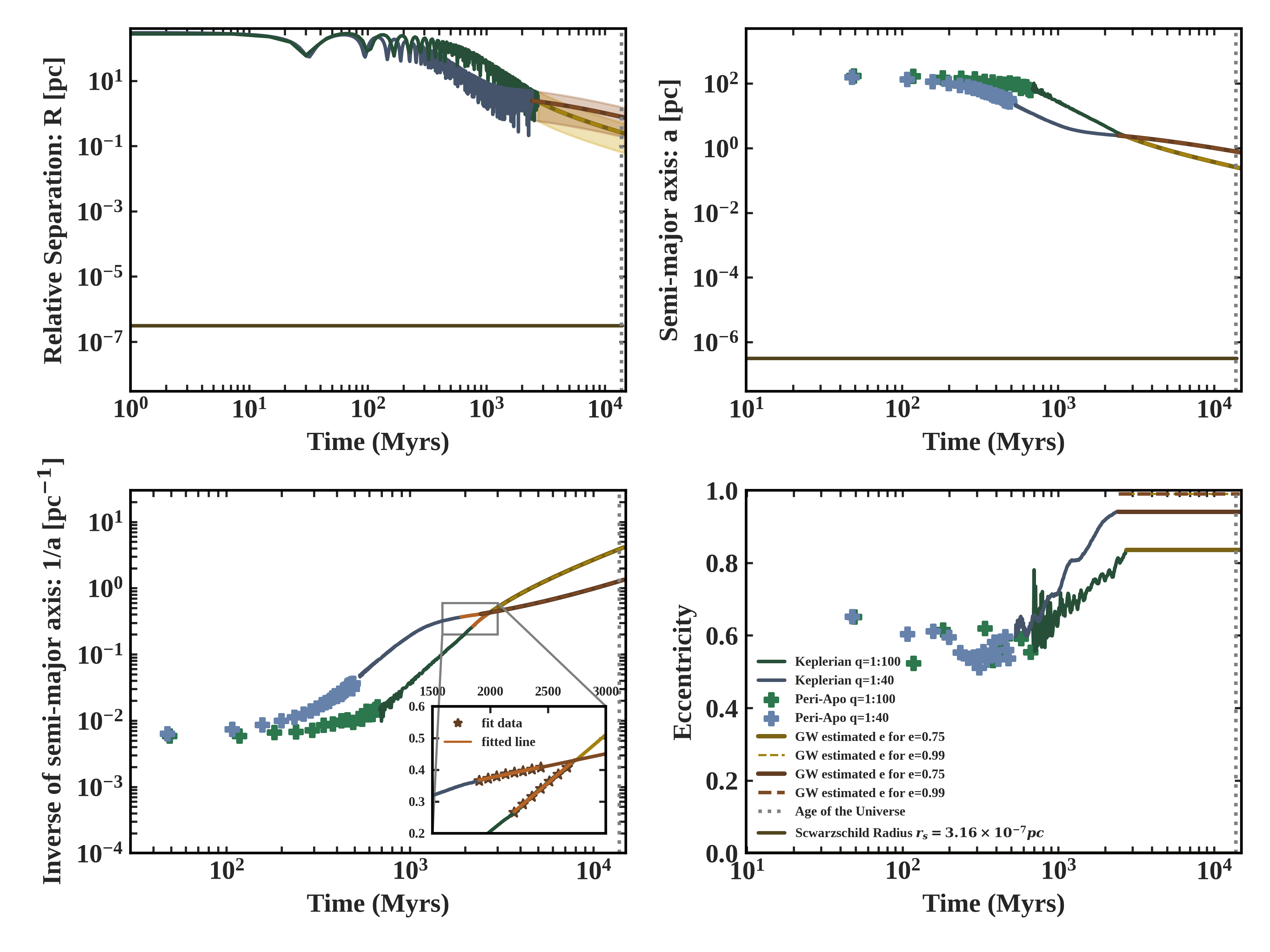}
        \caption{Orbital evolution of the massive black hole binary. Top-left: relative separation ($\Delta R$). Top-right: semi-major axis ($a$). Bottom-left: inverse semi-major axis ($1/a$), inside plot zooming to the region showing the slopes used to calculate hardening rates. Bottom-right: eccentricity ($e$). Time is measured from $t = 0$ when the secondary black hole reaches apoapsis at 300 pc. The plus markers denote values derived from apoapsis and periapsis estimates, while the corresponding solid lines represent values directly extracted from the simulation. Gravitational wave (GW) evolution is overplotted for initial eccentricities of $e_o = 0.75$ (bold solid line) and $e_o = 0.99$ (dashed line). Green and gold curves correspond to the $q = 0.01$ model, while blue and copper indicate the $q = 0.025$ model. The vertical grey dashed line marks the age of the Universe (13.8 Gyr), and the horizontal black line in the upper panels indicates the Schwarzschild radius of the primary black hole.  }
        \label{fig:orbital-panel}
    \end{figure*}

    \subsection{Binary Evolution Overview}

    Figure~\ref{fig:orbital-panel} presents a comprehensive view of the MBHB orbital evolution over 2.5 Gyr of simulation time for Leo I model.
    The upper-left panel shows the relative separation,  for $q = 0.01$, relative separatiion gradually shrinks from $\sim300$ pc to sub-parsec scales.
    The binary reaches $\sim$10 pc after $\sim$1.5 Gyr, and 5 pc near the 2 Gyr mark.
    But for $q = 0.025$ model, relative separation of BBH reaches 10 pc within 1 Gyr but after this the binary stalls, At no point does the separation drop below 0.1 pc, suggesting long-term stalling instead of moving towards gravitational-wave (GW) coalescence.
    The upper-right panel shows the semi-major axis $(a)$, which steadily declines following binary formation ($T_b \sim 0.5$ Gyr).
    Before this time the semi-major axis is estimated using the periapsis and apoapsis values for binary and later the solid line represents semi-major from the simulation's results.
    Similar to the relative separation, semi-major axis of the $q = 0.01$ system drops graually to the final parsec meanwhile $q = 0.025$ system reaches that mark quicker by almost 1 Gyr.
    The flattening trend beyond 1.5 Gyr suggests inefficient hardening.
    After that the semi-major axis in both simulations stalls from further dropping.
    In gravitational emission regime of BBH evolution, the average energy emitted in terms of GWs is given by~\cite{peters1964gravitational}:
    \begin{equation}
        \frac{dE_{rad}}{dt}=\frac{32}{5}\frac{G^4}{c^5}\frac{M_1^2M_2^2(M_1+M_2)}{a^5}F(e)
    \end{equation}
    where the factor
    \begin{equation}
        F(e)=(1-e^2)^{-7/2}\left(1+\frac{73}{24}e^2+\frac{37}{96}e^4\right)
    \end{equation}
    where $F(e)$ is a function dependent on the binary's eccentricity.
    So, the change in semi-major axis and eccentricity due to GW will be:

    \begin{equation}
        \left.\frac{da}{dt}\right|_\text{GW}=-\frac{64}{5}\frac{G^3}{c^5}\frac{M_1 M_2 (M_1 + M_2 )}{a^3} F(e)
    \end{equation}

    \begin{figure*}
        \centering
        \includegraphics[width=\textwidth]{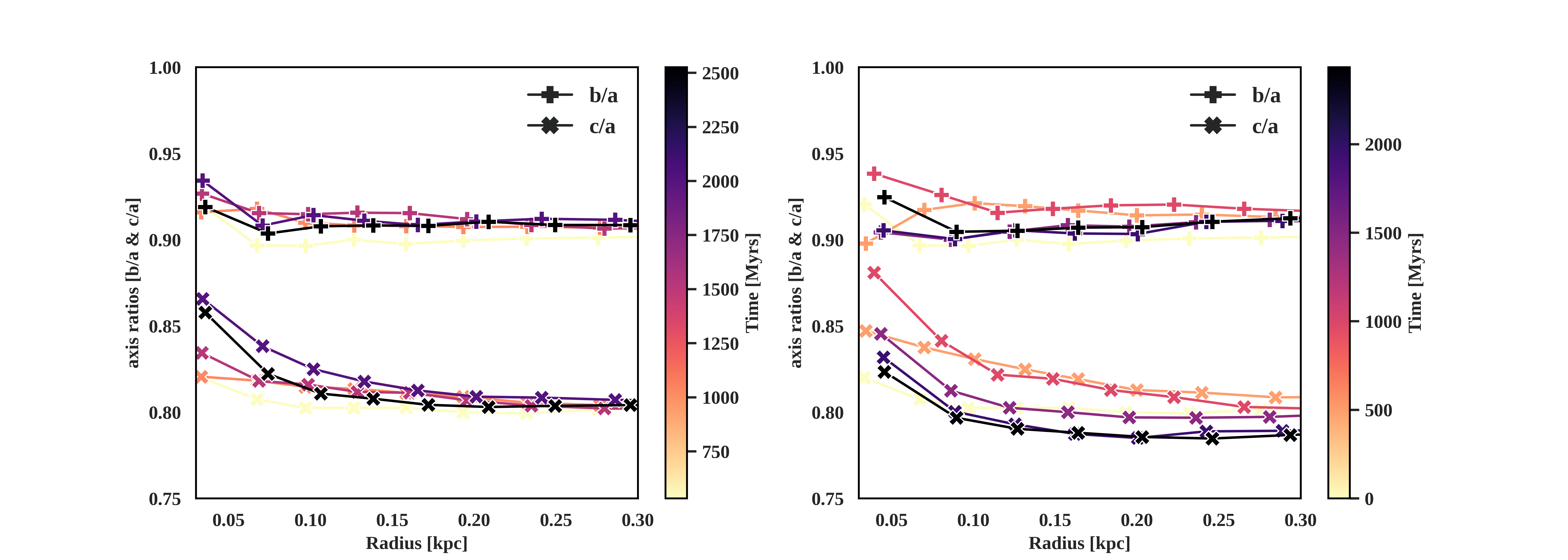}
        \caption{Time evolution of axis ratios in the $q=0.01$ model (left) and $q=0.025$ model  (right): $b/a$ as "$+$" and $c/a$ as "$\times$" and inner region (right). The system becomes more spherical over time, potentially limiting loss cone refilling.}
        \label{fig:triaxiality}
    \end{figure*}

    \begin{equation}
        \left.\frac{de}{dt}\right|_\text{GW} = -\frac{304}{15} \frac{G^3}{c^5}\frac{M_1 M_2 (M_1 + M_2 )}{a^4}  e(1-e^2)^{-5/2}\left(1+\frac{121}{304}e^2\right)
    \end{equation}

    So, the estimation of semi-major axis using the loss of orbital energy via GW suggests that the binaries will not merge by reaching the Schwarzchild radius of the primary black hole ($r_s$ = $3.16\times 10^{-7}$ pc) within the Hubble time as the change is very small over time.
    The bottom-left panel shows inverse semi-major axis ($1/a$), a useful parameter for analysis of binary hardening.
    Its decreasing slope implies progressive stalling.
    The hardening rate is given by:

    \begin{equation}
        s_{\mathrm{NB}}=\left(\frac{d}{dt}\left(\frac{1}{a} \right)\right)_{\mathrm{NB}} = -\frac{1}{a^2}\left(\frac{da}{dt}\right)_{\mathrm{NB}}
    \end{equation}

    \begin{equation}
        \left(\frac{da}{dt}\right)_{\mathrm{NB}} =  -s_{\mathrm{NB}}\cdot a^2, s_{\mathrm{NB}} \approx const
    \end{equation}

    \begin{equation}
        \left(\frac{da}{dt}\right)_{\mathrm{total}} =  -(s_{\mathrm{NB}}\cdot a^2) +  \left(\frac{da}{dt}\right)_{\mathrm{GW}}
    \end{equation}

    A hardening rate $s_{\rm NB} \approx 0.31$ kpc$^{-1}$ Myr$^{-1}$for  $q = 0.01$ and  $s_{\rm NB} \approx 0.076$ kpc$^{-1}$ Myr$^{-1}$ for $q = 0.025$ is inferred by the results of the simulation by calculating the slope of inverse semi-major axis around 2 to 2.5 Gyr.
    These extremely low values of hardening rates are then used to calculate semi-major axis and eccentricity till 14 Gyrs.
    We estimated semi-major axis and eccentricities from $a_o \approx 2.5$ pc and $e_o = 0.75$ for both models and as well as at extreme eccentricity, $e\approx 0.99$.
    The bottom-right panel displays the orbital eccentricity, estimated using apoapsis/periapsis extraction till the hard binary forms.
    Eccentricity grows gradually to $e \sim 0.85$ for $q = 0.01$ and $e \sim 0.95$ for $q = 0.025$ model by the end of the simulation.
    The greater value enhances GW energy loss but remains insufficient for coalescence within a Hubble time, given the stalled separation.

    \subsection{Axis ratio evolution}

    The initial shape of the galaxy model is nearly spherical with axis ratios $b/a = 0.9$, $c/a = 0.8$.
    Over the course of the simulation, the inner galaxy evolves toward a more spherical shape, with $c/a$ increasing to $\sim0.85$.
    This flattening trend, shown in Fig.~\ref{fig:triaxiality}, is likely induced by angular momentum redistribution from BBH–star interactions.
    A more spherical core may reduce the population of centrophilic orbits, suppressing further loss cone refilling.

    \subsection{Core Evolution}

\begin{figure*}
        \centering
        \includegraphics[width=\textwidth]{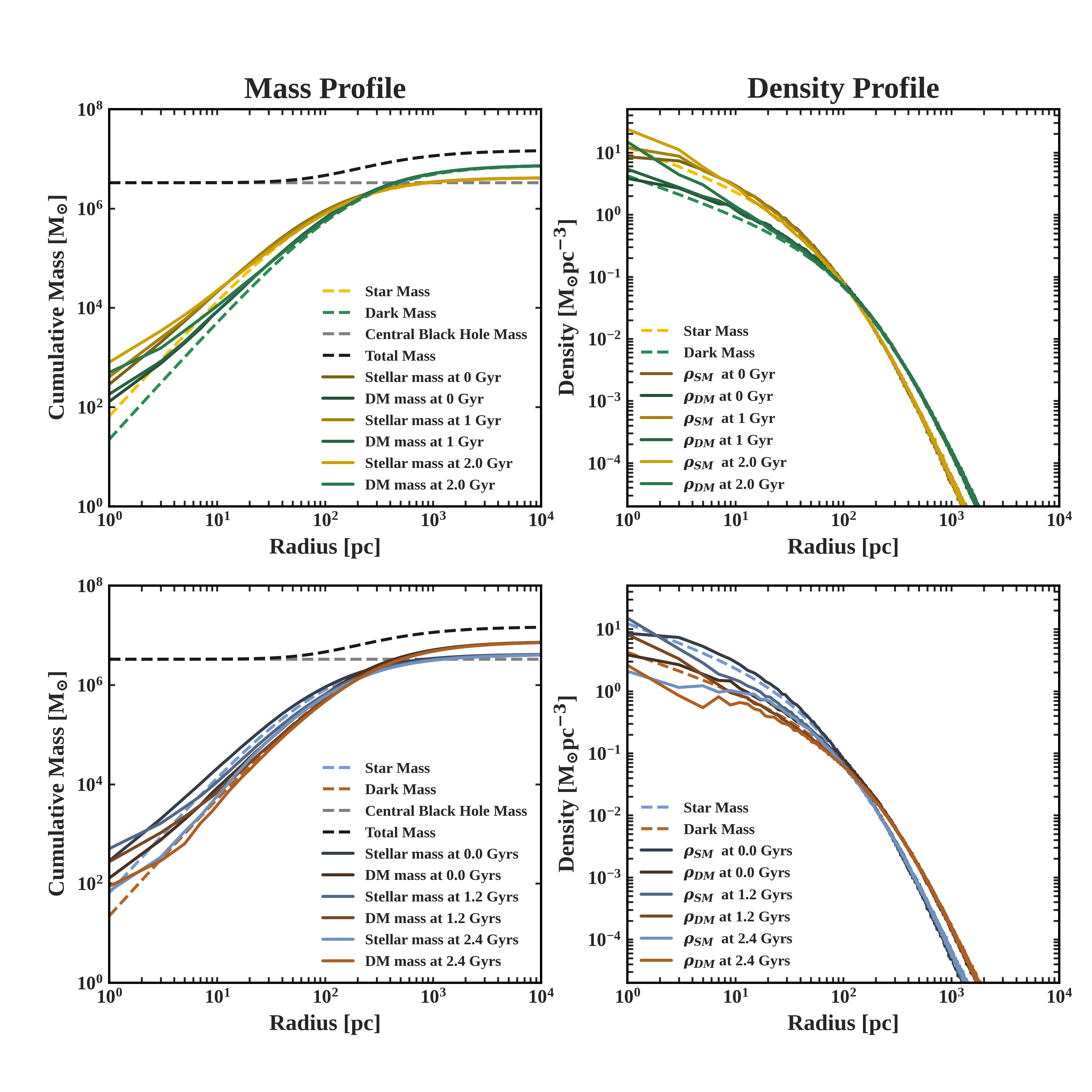}
        \caption{Evolution of mass and density profiles over 2.4 Gyrs.Top panels for $q_1$ and bottom panels for $q_2$ model.}
        \label{fig:stability}
    \end{figure*}

\begin{figure*}
        \centering
        \includegraphics[width=\textwidth]{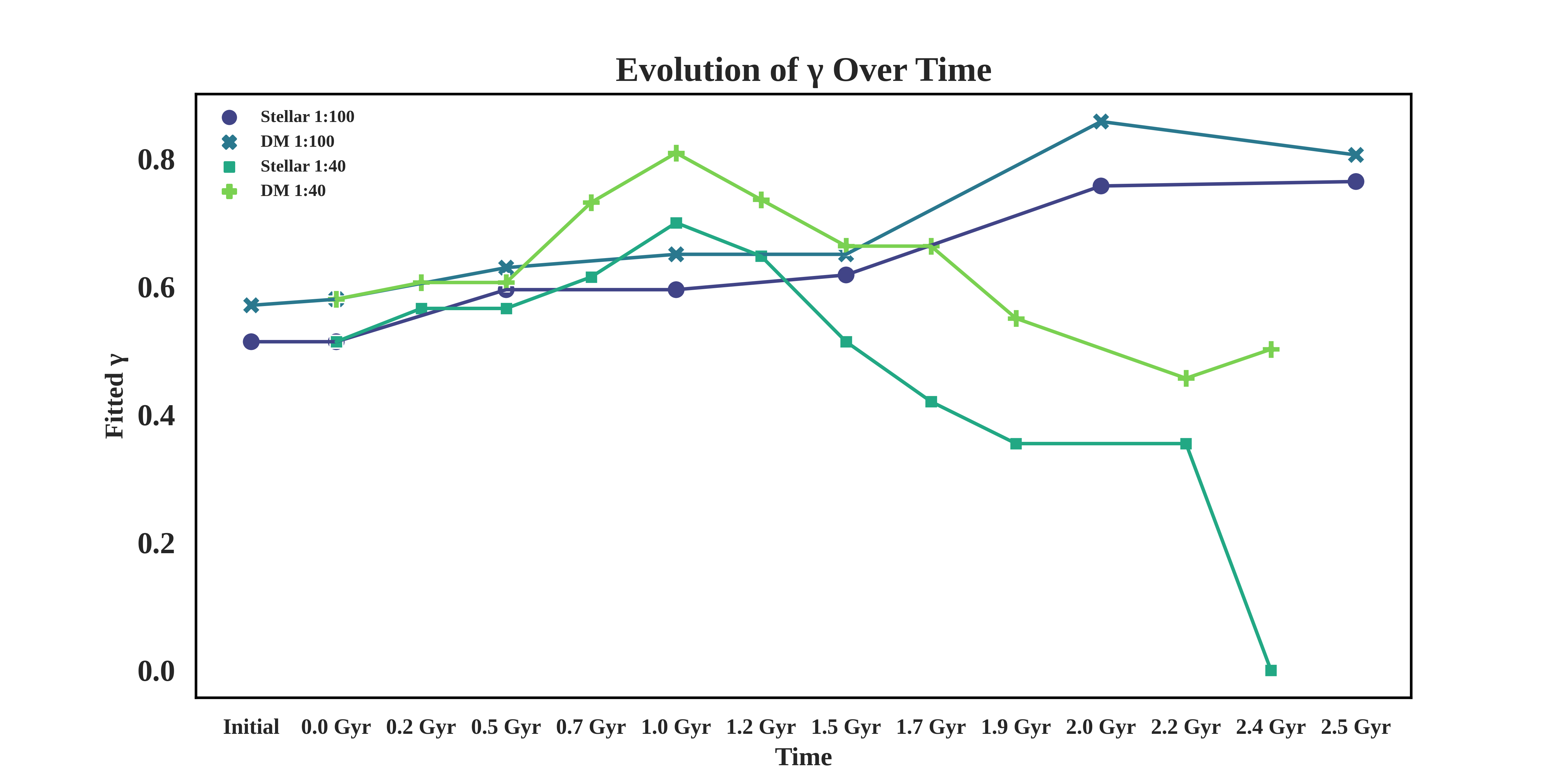}
        \caption{Mass and density evolution in the inner region. Profiles show central depletion due to BBH interactions, limiting further evolution.}
        \label{fig:inner-profile}
    \end{figure*}

    The MBHB’s hardening ejects stars via three-body encounters, depleting both stellar and dark matter densities in the central $\sim$50 pc.
    Figure~\ref{fig:stability} shows cumulative mass and density profiles at multiple timestamps.
    Over the 2 Gyr evolution, the core experiences a tenfold drop in density.
    The resulting mass deficit halts further hardening once the binary reaches $\sim$0.4 pc, leading to long-term stalling.
    Figure~\ref{fig:inner-profile} shows how gamma (inner slope) evolved from shallow cusp and initial value of 0.6 throughout the simulation period.
    For less massive MBHB model, it remains almost stable and ends at closer to 0.8, but for heavier MBHB, $\gamma$ first rises and peaks at almost 1 Gyr and then continues to drop and finally ends at 0.5 for dark matter and 0 or stellar matter, displaying core formation.

    \section{Conclusion and Discussion}\label{concluions}

    \subsection{Conclusion}

    We investigated the dynamical evolution of a massive black hole binary
    (MBHB) at the centre of a dwarf spheroidal galaxy, using Leo~I as a
    well-constrained observational template.
    A self-consistent, multi-component
    N-body model of Leo~I was constructed with \textsc{agama}~\cite{Vasiliev_2018}, incorporating the observed stellar and dark matter
    distributions together with the central black hole inferred from kinematic
    data~\cite{bustamante2021}.
    A stability analysis confirmed that the model
    preserves its initial density and mass profiles over 180 Myr, validating it
    as a reliable initial condition for binary evolution experiments.
    A secondary
    IMBH was then introduced at 300~pc with mass ratios $q_1 = 0.01$ and
    $q_2 = 0.025$ relative to the primary, and the subsequent evolution was
    followed for $\sim$2.5~Gyr using the direct N-body code $\varphi$-GPU~\cite{2011Berczik}.

    Our principal findings are as follows.

    \begin{enumerate}[label=(\roman*)]

    \item \textbf{Binary formation and dynamical friction.}  Both models
    form a gravitationally bound hard binary within $\sim$0.5--1~Gyr through
    dynamical friction against the stellar and dark matter background.
    The
    timescale is substantially longer than in more massive, denser systems,
    a direct consequence of Leo~I's comparatively low stellar and dark matter
    densities, which are themselves a legacy of repeated tidal stripping by
    the Milky Way~\cite{pacucci2023extreme}.

    \item \textbf{Hardening and the final parsec problem.}  Following binary
    formation, three-body stellar encounters drive the binary separation from
    $\sim$300~pc to sub-parsec scales.
    The $q = 0.025$ model reaches the
    final parsec somewhat faster than the $q = 0.01$ model, but both
    ultimately stall at separations of order $\sim$0.4--1~pc.
    The inferred
    hardening rates are $s_{\rm NB} \approx 0.31$~kpc$^{-1}$~Myr$^{-1}$
    for $q = 0.01$ and $s_{\rm NB} \approx 0.076$~kpc$^{-1}$~Myr$^{-1}$
    for $q = 0.025$, among the lowest reported in binary evolution studies,
    reflecting the depleted loss cone in this low-density environment.

    \item \textbf{Gravitational-wave evolution.}  Extrapolating beyond the
    simulated period using Peters' equations~\cite{peters1964gravitational},
    neither model merges within a Hubble time.
    Even for the extreme case of
    $e = 0.99$, the binary only contracts to $\sim$0.4~pc over 13.8~Gyr,
    falling far short of the Schwarzschild radius of the primary
    ($r_s = 3.16\times10^{-7}$~pc).

    \item \textbf{Eccentricity growth.}  Orbital eccentricity grows
    progressively during the hardening phase, reaching $e \sim 0.85$ for
    $q = 0.01$ and $e \sim 0.95$ for $q = 0.025$.
    Although higher
    eccentricity enhances gravitational-wave energy loss, this enhancement is
    insufficient to overcome the stalled separation in the absence of an
    efficient loss-cone replenishment mechanism.

    \item \textbf{Inner morphological evolution.}  Three-body ejections from
    the binary heat and scatter the surrounding stellar and dark matter
    distributions, driving the inner morphology toward a more spherical
    configuration ($c/a$ increases from $\sim$0.8 to $\sim$0.85) and
    reducing the population of centrophilic orbits available for loss-cone
    refilling.
    Simultaneously, the central density drops by roughly an
    order of magnitude within 50~pc, and the inner power-law slope $\gamma$
    shows model-dependent behaviour: it remains near 0.8 for the less
    massive secondary but first rises then falls below 0.5 for the more
    massive secondary, indicating the onset of core formation.

    \end{enumerate}

    Taken together, these results demonstrate that while a SMBH--IMBH binary
    can assemble dynamically in a Leo~I-type dwarf spheroidal, the final parsec
    problem persists robustly.
    The low-density, tidally stripped environment
    of Leo~I is insufficient to drive the binary to coalescence through stellar
    dynamics alone, even when dark matter is included.
    Additional physical
    processes -- such as gas inflows, residual triaxiality, or the presence of
    a nuclear star cluster -- would be needed to overcome this barrier.

    \subsection{Discussion}

    The final parsec problem has long been recognised as a potential bottleneck
    in the merger history of massive black holes~\cite{2015Sesana_Khan}, and our
    results provide a concrete illustration of how acutely it manifests in the
    low-density environment of a dwarf spheroidal.
    In more massive, gas-rich
    systems, AGN feedback and nuclear gas flows can replenish the loss cone on
    relatively short timescales.
    In Leo~I, however, the stellar mass
    is comparatively low and largely tidally depleted~\cite{pacucci2023extreme},
    and no significant gas reservoir is expected, leaving three-body stellar
    scattering as the dominant hardening channel -- one that ultimately proves
    insufficient.

    The question of whether the Leo~I black hole is a descendant of a
    heavy seed formation event has been raised by recent theoretical work.
    \citet{scoggins2025heavyblackholeseed} argue that a black hole of
    $\sim 3 \times 10^6~\mathrm{M_\odot}$ residing in a system with Leo~I's
    properties is broadly consistent with the expected descendants of heavy
    black hole seeds, motivating the interpretation of Leo~I as a surviving
    relic of an early seed formation site.
    If this picture is correct, the
    binary studied here can be thought of as the dynamical encounter between
    a heavy seed survivor and an infalling IMBH. We note, however, that the
    existence and mass of the Leo~I black hole itself remains debated; some
    studies argue that the observed kinematic signal may be reproduced by an
    unusually high dark matter concentration in the nucleus, without requiring a
    black hole at all~\cite{scoggins2025heavyblackholeseed}.
    The conclusions
    drawn here are therefore conditional on the black hole being real.

    A related astrophysical channel involves Leo~I itself being accreted by the
    Milky Way.
    Dwarf galaxies like Leo~I are expected to lose orbital energy
    to dynamical friction and eventually sink toward the centres of their more
    massive hosts.
    Should this occur, the Leo~I black hole could be delivered
    to the Galactic nucleus and pair with the Milky Way's central black hole,
    Sgr~A$^*$, forming what is termed a heavy IMRI -- an intermediate-mass ratio
    inspiral between a supermassive and an intermediate-mass black hole~\cite{askar2024intermediatemassblackholesstar}.
    Dense stellar clusters
    carrying IMBHs can undergo a similar infall process, reaching galactic
    nuclei on timescales that depend on their birth radii and host properties~\cite{askar2024intermediatemassblackholesstar}.
    Within Leo~I itself, however, our simulations suggest that the SMBH--IMBH
    binary formed internally will not reach the gravitational-wave band within
    a Hubble time.
    This places Leo~I-type dSphs in a different category from dwarf galaxies with nuclear star clusters, where higher central densities can sustain loss-cone refilling and drive binaries toward coalescence on more favourable timescales~\citep{2021Khan_Holley-Bockelmann}.

    More broadly, these results highlight the sensitivity of MBHB evolution to the host galaxy environment.
    Dwarf spheroidals occupy the low end of the galaxy mass function~\cite{Mezcua_2017}, and their black hole populations remain poorly characterised observationally.
    Understanding whether and how
    binaries in such systems can merge is important for constructing realistic
    LISA source populations and for interpreting the nanohertz gravitational
    wave background detected by pulsar timing arrays~\cite{2017Amaro-Seoane}.
    Our work suggests that isolated, gas-poor dwarf spheroidals are unlikely to
    be significant contributors to either signal, but that the same black holes
    may become relevant IMRI sources if and when their host galaxies are
    absorbed into more massive systems.

    \section*{Acknowledgments}

    The author gratefully acknowledges Vanderbilt University’s Advanced Computing Center for Research and Education (ACCRE) and the National Center of GIS and Space Application (NCGSA)'s Space and Astrophysics Research Lab (SARL) and Modeling and Simulation Lab for providing the computational resources used for this research.

    \section*{Data Availability Statement}

    The data underlying this article will be shared on reasonable request to the corresponding author.



    \bibliographystyle{mnras}
    \bibliography{ms}

    \bsp    
    \label{lastpage}

\end{document}